\documentclass[conference]{IEEEtran}
\usepackage{tikz}
\usepackage{comment}

\newcommand\copyrighttext{%
  \footnotesize \textcopyright 2024 IEEE. Personal use of this material is permitted. Permission from
IEEE must be obtained for all other uses, in any current or future media,
including reprinting/republishing this material for advertising or promotional
purposes, creating new collective works, for resale or redistribution to servers
or lists, or reuse of any copyrighted component of this work in other works.}
\newcommand\copyrightnotice{%
\begin{tikzpicture}[remember picture,overlay]
\node[anchor=south,yshift=10pt] at (current page.south) {{\parbox{\dimexpr\textwidth-\fboxsep-\fboxrule\relax}{\copyrighttext}}};
\end{tikzpicture}%
}

\usepackage{cite}
\usepackage{amsmath,amssymb,amsfonts}
\usepackage{algorithmic}
\usepackage{graphicx}
\usepackage{textcomp}
\usepackage{xcolor}

\renewcommand\IEEEkeywordsname{Key words}
\newcommand{\linebreakand}{%
  \end{@IEEEauthorhalign}
  \hfill\mbox{}\par
  \mbox{}\hfill\begin{@IEEEauthorhalign}
}
\def\BibTeX{{\rm B\kern-.05em{\sc i\kern-.025em b}\kern-.08em
    T\kern-.1667em\lower.7ex\hbox{E}\kern-.125emX}}
    
\begin{document}

\title{Stability in Isolated Grids: Implementation and Analysis of the Dead-Zone Virtual Oscillator Control in Simulink and Typhoon HIL \\
}

{\author{\IEEEauthorblockN{Dixant Bikal Sapkota}
\IEEEauthorblockA{\textit{Department of Electrical Engineering} \\
\textit{IOE Pulchowk, Tribhuvan University}\\
Lalitpur, Nepal \\
dixantbs7@gmail.com}
\and
\IEEEauthorblockN{Puskar Neupane}
\IEEEauthorblockA{\textit{Department of Electrical Engineering} \\
\textit{IOE Pulchowk, Tribhuvan University}\\
Lalitpur, Nepal\\
pneupane2024@gmail.com}
\and
\IEEEauthorblockN{Bivek Shiwakoti}
\IEEEauthorblockA{\textit{Department of Electrical Engineering} \\
\textit{IOE Pulchowk, Tribhuvan University}\\
Lalitpur, Nepal \\
bivekski1@gmail.com}

\and
\IEEEauthorblockN{Saugat Baral}
\IEEEauthorblockA{\textit{Department of Electrical Engineering} \\
\textit{IOE Pulchowk, Tribhuvan University}\\
Lalitpur, Nepal\\
99saugat@gmail.com}

\and
\IEEEauthorblockN{Panas Bhattarai}
\IEEEauthorblockA{\textit{Department of Electrical Engineering} \\
\textit{IOE Pulchowk, Tribhuvan University}\\
Lalitpur, Nepal \\
acpanasbhattarai@gmail.com}
\and

\IEEEauthorblockN{Dr. Basanta Kumar Gautam}
\IEEEauthorblockA{\textit{Department of Electrical Engineering} \\
\textit{IOE Pulchowk, Tribhuvan University}\\
Lalitpur, Nepal \\
basanta.gautam@pcampus.edu.np}

}
}
\maketitle

\copyrightnotice

\begin{abstract}
This paper explores the analysis and implementation of the Dead-Zone Virtual Oscillator Control (DZVOC) strategy for grid-forming inverters aiming to enhance stability amidst the ever-increasing penetration of renewable energy sources like photo-voltaic and wind. Key objectives include implementation and analysis of a DZVOC three-phase battery-inverter system with a voltage control loop on top, study of its stability and performance in an isolated micro-grid and exploration of their use alongside a grid following PV-inverter system with a hysteresis band current control. By modelling independent microgrids under various cases such as VOC inverters of varying capacities and a VOC inverter in conjunction with a PV inverter, this research addresses critical aspects of power-sharing, compatibility, response times, and fault ride-through potential, as well as improving the voltage droop profile of a general DZVOC control. The simulation is executed in Matlab Simulink and validated with real-time simulation using the Typhoon-HIL 404.
\end{abstract}

\begin{IEEEkeywords}
Grid Forming, Virtual Oscillator Control, Hysteresis band, Hardware-in-the-loop
\end{IEEEkeywords}

\section{Introduction}
The need for stable and reliable control for Inverter Based Resources (IBRs) is immense with the increasing penetration of renewable energy sources, like solar and wind. The nature of traditional sources, consisting of large rotating machines, has a unique way of providing stability to the system through the inertia of their mass. This ensures that the interconnected power system can operate reliably, allowing control systems to account for disturbances. However, this phenomenon cannot be leveraged by inverter-based sources. Solar PV arrays cannot provide any mechanical inertia and wind turbines have intermittency in their rotations requiring AC-DC-AC conversion or other complex control systems. All this points to the fact that the growing share of IBRs causing a decrease in inertia in the system may lead to issues regarding the stability and reliability of the system. 

Over the decades, major research has been conducted to develop the field of Grid Forming Inverters (GFMs). GFM inverters act as voltage sources and can stabilize their frequency, thus having the capability to operate in isolation. Virtual Synchronous Machines (VSM) and Droop control are two of the common methodologies. Both control logics imitate the physics of a real alternator, allowing an inverter to behave in a very similar fashion. These methods introduce virtual inertia to the system. While they have their strengths, emulating a complex physical system comes with its challenges. It requires high computational power, is prone to convergence issues and can introduce time delays due to the calibre of calculations. 

Synchronization of complex networks through oscillator models have been studied in \cite{Synchronization_of_oscillators_in_LTI, Sync_of_complex_dynamical_networks, Sync_of_coupled_oscillator_dynamics} emphasizing that synchronization is a very significant phenomenon to study the collective behaviour of coupled oscillators similar to that in an interconnected power system. Based on these studies, a newer grid forming control, the Virtual Oscillator Control (VOC) \cite{Voltage_control_for_voltage_source_inverters, Synchronization_of_parallel_single_phase_VOC}, has grasped the attention of many due to its very unique characteristics. This control logic has: quick response times with inherent and accurate power sharing between any number of inverters, less computational burden with simple and straightforward design, and requires no explicit power calculations or external references \cite{VOC_review_paper_Agamy}. Sufficient conditions for the global synchronization of identical nonlinear oscillators have been derived.

The comparative analysis of VOC with other control methodologies has been carried out by \cite{Comparison_droop_vsm_voc_isolated_microgrid, Dhople_Comparison_VOC_Droop, Droop_and_Andron_Hopf, Three_Phase_VOC_Droop, VOC_Droop_Frontiers, voc_compare_PLL_frame}. While all control methods ensure synchronization and power-sharing, VOC is touted as the quickest as it acts on instantaneous measurements. Also, various control architectures exist within the scope of VOCs, which have been studied by \cite{Dispatchable_VOC, Dhople_AHO_grid_compatible, Comparison_of_DZ_VDP_Microgrid,Adaptation_of_current_control_to_voc,VOC_Ajso}, which discuss common controls like Van der Pol Oscillator, Dead-Zone Oscillator, dispatchable-VOC, and Andronov-Hopf Oscillator as well as adaptations of general current controlled control systems to incorporate oscillator controls.

The remainder of this paper is organized as: Section $II$ contains the formulation, analysis, and system diagram for dead-zone virtual oscillator control, standalone micro-grid under study, and PV-inverter system. Section $III$ is the result containing graphs which show the responses of the system in various conditions. Section $IV$ concludes the paper.

\section{Methodology}
\subsection{Virtual Oscillator Control}
\subsubsection{Dead Zone Oscillator}

\begin{figure}[!b]
    \centering
    \includegraphics[width=0.48\textwidth]{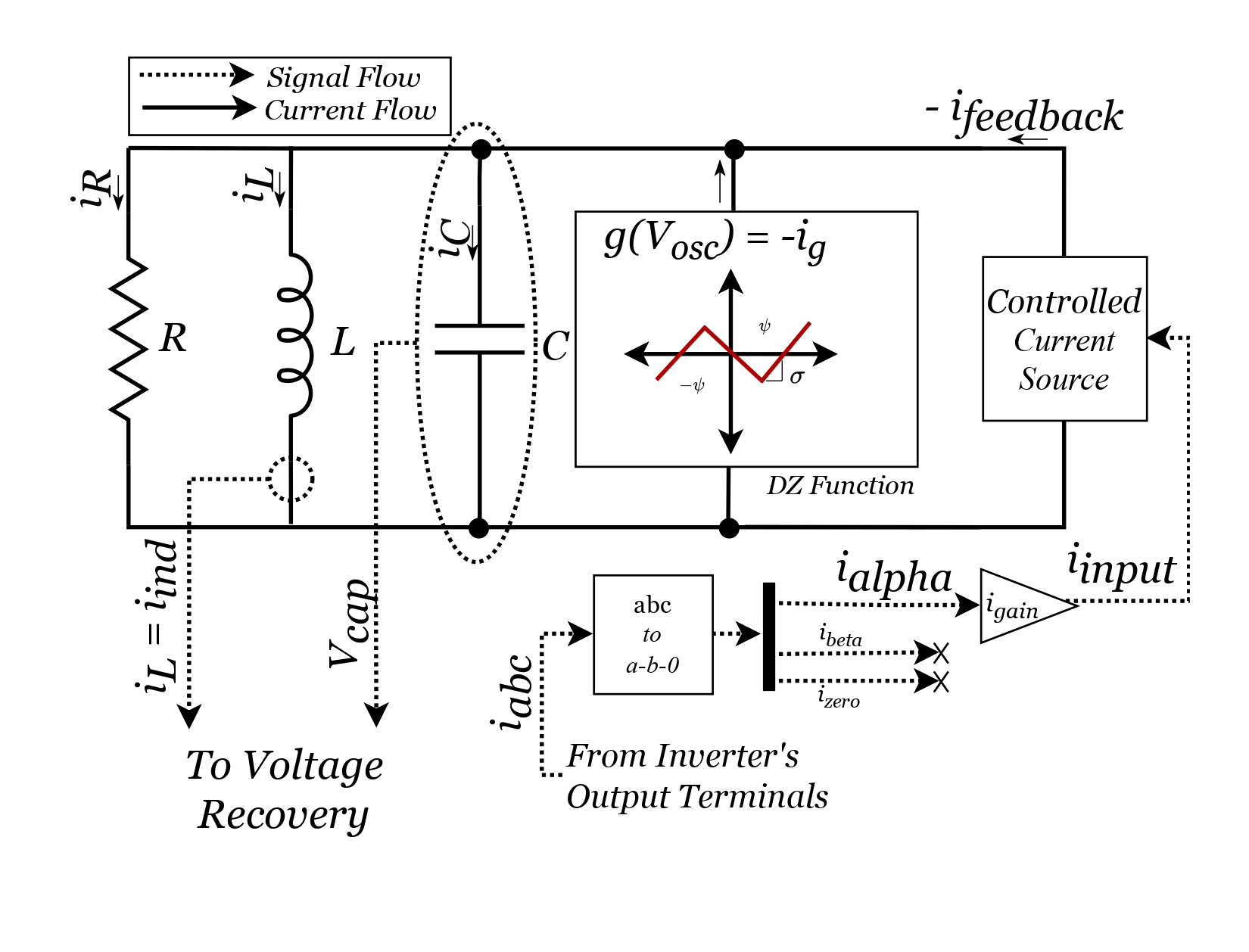}
    \caption{Circuit Model of Dead Zone Oscillator }
    \label{DZO_Circuit}
\end{figure}

The Dead Zone Oscillator (DZO) is a type of the VOC strategy, which is
characterized by a piece-wise non-linear function \cite{VOC_SOTA_Review}. This function, as given in Equation \ref{dz_eqn}, is implemented which outputs current to the oscillator circuit based on the instantaneous voltage across it. The circuit model of a three-phase dead zone oscillator is shown in Figure \ref{DZO_Circuit}.

\begin{equation}
\label{dz_eqn}
 f(V_{osc})=
    \left\{
    \begin {aligned}
         & 2\sigma(V_{osc}-\phi), \quad  V_{osc} > \phi \\
          & 0, \quad  \quad \quad \quad \quad \quad \left| V_{osc} \right| \le \phi \\
         & 2\sigma(V_{osc}+\phi), \quad  V_{osc} < -\phi \\
    \end{aligned}
     \right.
\end{equation}

where $V_{osc}$ is the DZO terminal voltage. In the given equation $2\sigma$ is the maximum slope of the voltage-dependent current source function $g(V_{osc})$ which is given in Equation \ref{g_function}.

\begin{equation}
\label{g_function}
    g(V_{osc}) = f(V_{osc}) - \sigma V_{osc}
\end{equation}

where, $g(V_{osc})$ represents the output current from the dead-zone function block to the oscillator. The transfer function of the impedance circuit as seen in Figure \ref{DZO_Circuit} is given by Equation \ref{RLC_transfer}.

\begin{equation}
    \label{RLC_transfer}
    z_{osc}(s) = (R || sL ||(sC)^{-1}) = \frac{\frac{1}{C}s}{s^{2}+\frac{1}{RC}s + \frac{1}{LC}}
\end{equation}

Considering the relation in the first condition of Equation \ref{dz_eqn},

\begin{equation}
  \begin{aligned}
  i_{g} & = -(2\sigma(v_{osc}-\phi)-\sigma v_{osc});\;\\
  & = -(\sigma(v_{osc}-2\sigma \phi)
  \end{aligned}
\end{equation}

The voltage appearing across the oscillator circuit, $V_{osc}$, is essentially the voltage across the capacitor. As a parallel circuit, the same voltage appears across the inductor terminals. The relation between current and voltage for an inductor is given by Equation \ref{inductor_current_eqn}.

\begin{equation}
\label{inductor_current_eqn}
i_{L} = \frac{1}{L}\int v_{osc} \; dt
\end{equation}

If $i_{abc}$ is the output current at the inverter terminals, then,
\begin{equation}\nonumber
i_{feedback} = i_{gain}\times i_{alpha} = i_{gain}\times \frac{V_{alpha}}{R_{s}} = i_{gain}\times v_{gain}\frac{v_{osc}}{R_{s}}
\end{equation}

\begin{center}
    where, $\alpha=i_{gain}\times v_{gain} =$ $overall$ $gain$,\\
    \vspace{0.1cm}
\end{center}
$v_{osc}$ is the $alpha$ component during reference generation for the Pulse Width Modulator (PWM). From Clarke's transform,

\begin{center}
    \vspace{0.1cm}
    $i_{alpha} = \sqrt(\frac{2}{3}) (i_{a} -\frac{i_{b}}{2} - \frac{i_{c}}{2})$\\
    \vspace{0.2cm}
    $v_{alpha} = \sqrt(\frac{2}{3}) (v_{a} -\frac{v_{b}}{2} - \frac{v_{c}}{2})$
\end{center}

Applying Kirchoff's current law at the junction point of the RLC block and dead zone function block, we obtain Equation \ref{kcl_equation}.

\begin{equation}
\label{kcl_equation}
i_{c} = \frac{\partial v_{osc}}{\partial t} = \frac{-v_{osc}}{R}+i_{g}-i_{L}-i_{feedback}
\end{equation}

\begin{center}
$or,\frac{\partial v_{osc}}{\partial t} = \frac{1}{C}[(\rho)v_{osc}-(2\sigma(v_{osc}-\phi))-\frac{1}{L}\int v_{osc}\;dt-\alpha\frac{v_{osc}}{R_{s}}$

\vspace{0.1cm}

{where, $i_{g} = -g(v_{osc}) = -[f(v_{osc})-\sigma v_{osc}$]}
\vspace{0.2cm}\\
{and $\rho = \sigma - \frac{1}{R}$}
\end{center}

Further differentiating and solving,

\begin{equation}\nonumber
\frac{\partial^2 v_{osc}}{\partial t^2} = \frac{1}{C}[\rho-2\sigma-\frac{\alpha}{R_{s}}]\frac{\partial V_{osc}}{\partial t}-\frac{1}{LC}v_{osc}
\end{equation}

\begin{equation}\nonumber
Let, \enspace \beta = \rho-2\sigma-\frac{\alpha}{R_{s}}, \enspace x_{o} = v_{osc},\enspace and,\enspace x_{1} = \frac{\partial V_{osc}}{\partial t}
\end{equation}

Here $x_{o}$ and $x_{1}$ are arbitrary variables.
Then,

\begin{equation}\nonumber
\frac{\partial^2 v_{osc}}{\partial t^2} = \frac{\beta}{C}\frac{\partial V_{osc}}{\partial t}-\frac{1}{LC}v_{osc}
\end{equation}

And,
\begin{equation}
\dot{x_{1}} = \frac{\beta}{C}x_{1}-\frac{1}{LC}x_{o}
\end{equation}
\vspace{-0.2cm}

So,
\begin{center}$\begin{bmatrix}\dot{x_{o}} \\\dot{x_{1}} \end{bmatrix} = \begin{bmatrix}0 & 1 \\-\frac{1}{LC} & \frac{\beta}{C} \end{bmatrix}\begin{bmatrix}x_{o} \\x_{1} \end{bmatrix}$\end{center}

Let $\lambda$ be the eigen values. Then,

\begin{center}$\begin{vmatrix}-\lambda & 1 \\-\frac{1}{LC} & \frac{\beta}{C}-\lambda \end{vmatrix} = 0$\end{center}

Which gives,
\vspace{-0.5cm}
\begin{equation}
\label{stability_eqn}
\lambda^{2}-\lambda(\frac{\beta}{C})+\frac{1}{LC} = 0
\end{equation}

\begin{figure}[!t]
    \centering
    \includegraphics[width=0.48\textwidth]{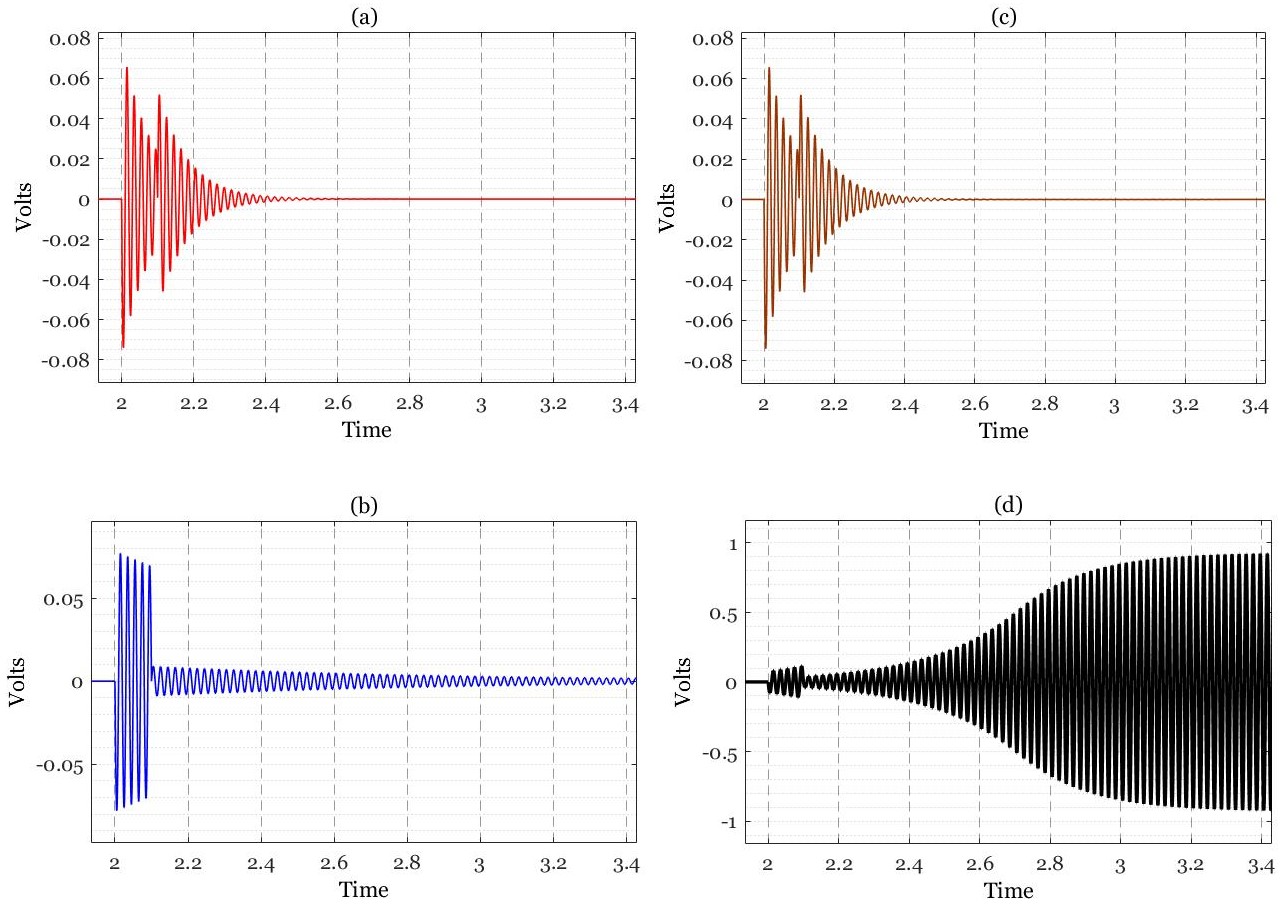}
    \caption{Impulse response of the oscillator circuit considering different parts of the dead-zone piece-wise function ($f(V_{osc})$): (a): $V_{osc} > \psi$, (b): $|V_{osc}| \le \psi$, (c): $V_{osc} < -\psi$, (d): $f(V_{osc})$ wholly.}
    \label{impulse_response}
\end{figure}

Solving Equation \ref{stability_eqn} using parameters listed in Table \ref{parameter_table} gives $\lambda = -7.3\pm j100\pi $ when $R_{s} = R$ is taken. Through substitutions, similar equations for other parts of the dead-zone function can be determined. If $\lambda_{ref} = a+jb$ for any system, it is stable as long as $a < 0$, and $b = 100\pi$ represents 50 Hz oscillations. This is visualized by Figure \ref{impulse_response} (a-c) where the output slowly dies out for each part of the piece-wise function separately. When the three linear equations are combined to form the entirety of the dead-zone nonlinear function, it shows sustained oscillations observed in Figure \ref{impulse_response}(d).

The inductance (L) and capacitance (C) of the oscillator are chosen to maintain 50Hz oscillations for our particular system. Resistance of the oscillator is selected as per the criteria: $\sigma > \frac{1}{R}$, to maintain a unit-circular phase plot which represents a sinusoidal wave in the time domain \cite{Synchronization_of_parallel_single_phase_VOC}. Filter impedances are chosen as a trade-off between excessive voltage drop and harmonic generation in the system. The gain values selected correspond to the voltage level and power capacity requirements.

\begin{figure}[!b]
    \centering
    \includegraphics[width=0.35\textwidth]{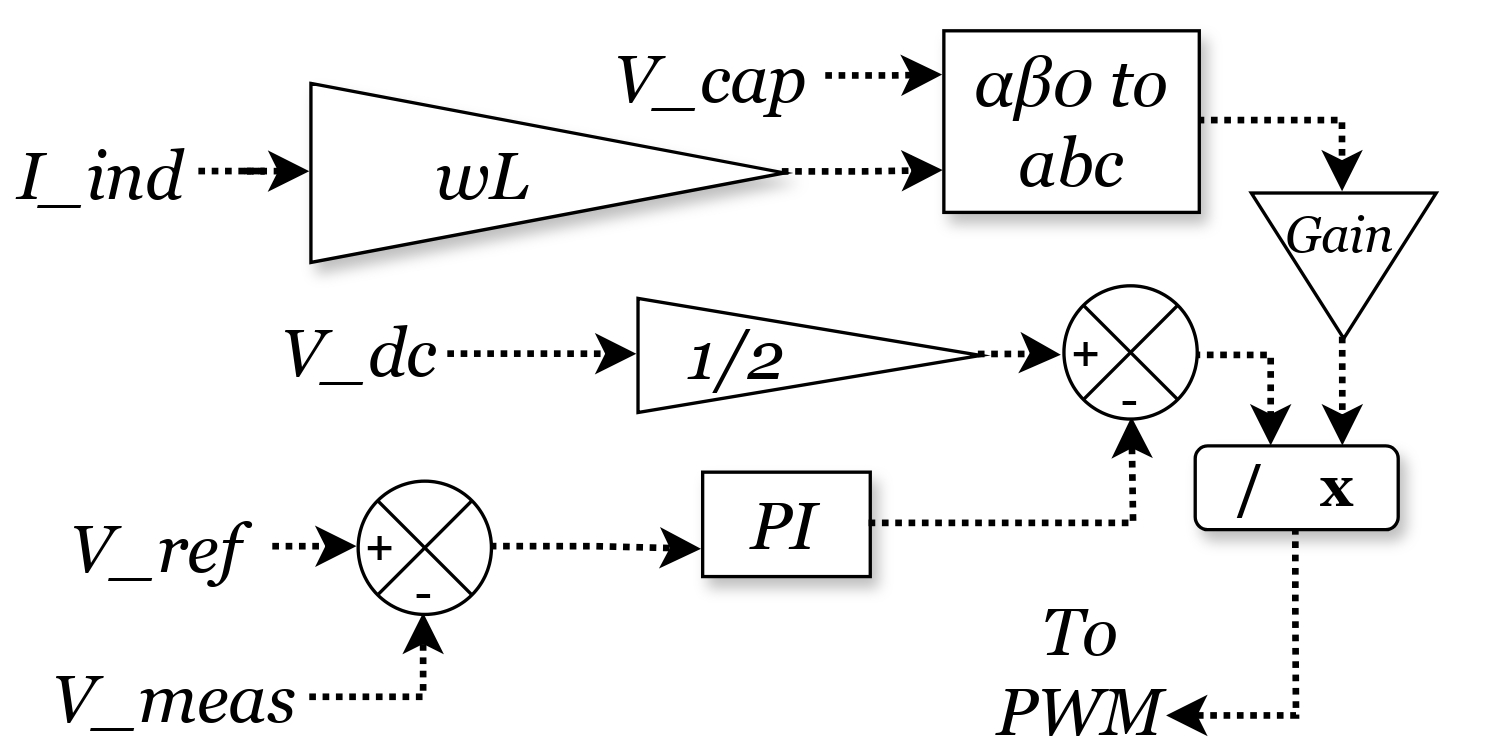}
    \caption{Voltage Recovery Loop (VRL)}
    \label{voltage_recovery_loop}
\end{figure}

The output current, at the inverter terminal, is measured and fed back to the control circuit through Clarke transformation. The first component $i_{alpha}$ induces a voltage across the R-L-C circuit. A Dead-zone function adjusts the current in the loop, based on this voltage. Just before the PWM generator, an inverse Clarke transformation is performed with two components: voltage across the capacitor ($V_{cap}$) and $w\cdot L$ as alpha and beta components respectively. An additional feature of the VOC control is its inherent power-sharing capacity. This is achieved by adjusting the gain parameters of the control loop in required multiples. In a standalone grid, filter parameters have to be adjusted in tandem with the current gain to counter the voltage drops across it. 

A Voltage Recovery Loop (VRL) has been added on top of the VOC control, as seen in Figure \ref{voltage_recovery_loop}. This loop adjusts the voltage at the output terminal of the inverter and uses a PI controller to adjust the gain of the reference voltage to the PWM, without obstructing the fast response of VOC itself. 

\begin{table}
\centering
\caption{Parameter selection}
\label{parameter_table}
\begin{tabular}{|c|c|c|c|}
\hline
Parameter & Value & Parameter & Value \\ \hline
$R$ & $10\Omega$ & $L_{f}$ &  $250e-4H$\\ 
$L$ & $250e-6H$ & $C_{f}$ &  $60e-6F$\\ 
$C$& $0.04052847F$ & $i_{gain}$ &$1.0568e^{-3}$ \\ 
$R_{f}$ &  $0.1\Omega$ & $v_{gain}$ & $(400*\sqrt(2))/\sqrt(3)$\\ 
\hline
\end{tabular}
\end{table}

\subsubsection{Multiple VOCs in a microgrid}
\begin{figure}[!b]
    \centering
    \includegraphics[width=0.5\textwidth]{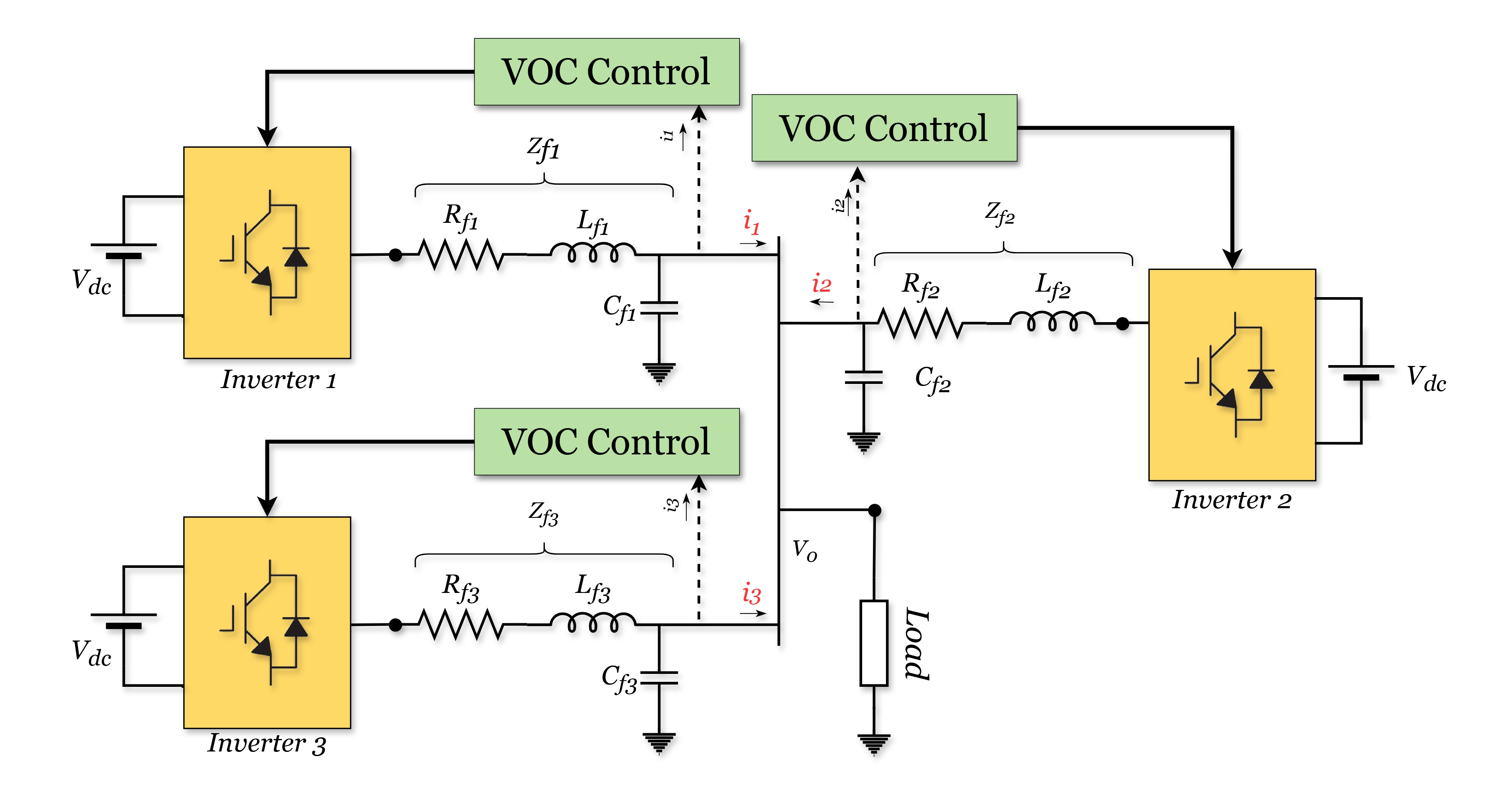}
    \caption{Standalone three-phase grid with differently sized inverter-based sources. }
    \label{voc_grid}
\end{figure}

Three different battery-inverter-based sources supply a common load together in an isolated three-phase micro-grid as seen in Figure \ref{voc_grid}. Each of them is controlled via DZ-VOC with VRL. The size (capacity) of each source is determined by the control parameters of each inverter. In this system, inverter 2 and inverter 3 are half and one-third capacity, respectively, of inverter 1. Each inverter takes the current at its terminal as the feedback.

\begin{figure}[!b]
    \centering
    \fbox{\includegraphics[width=0.45\textwidth]{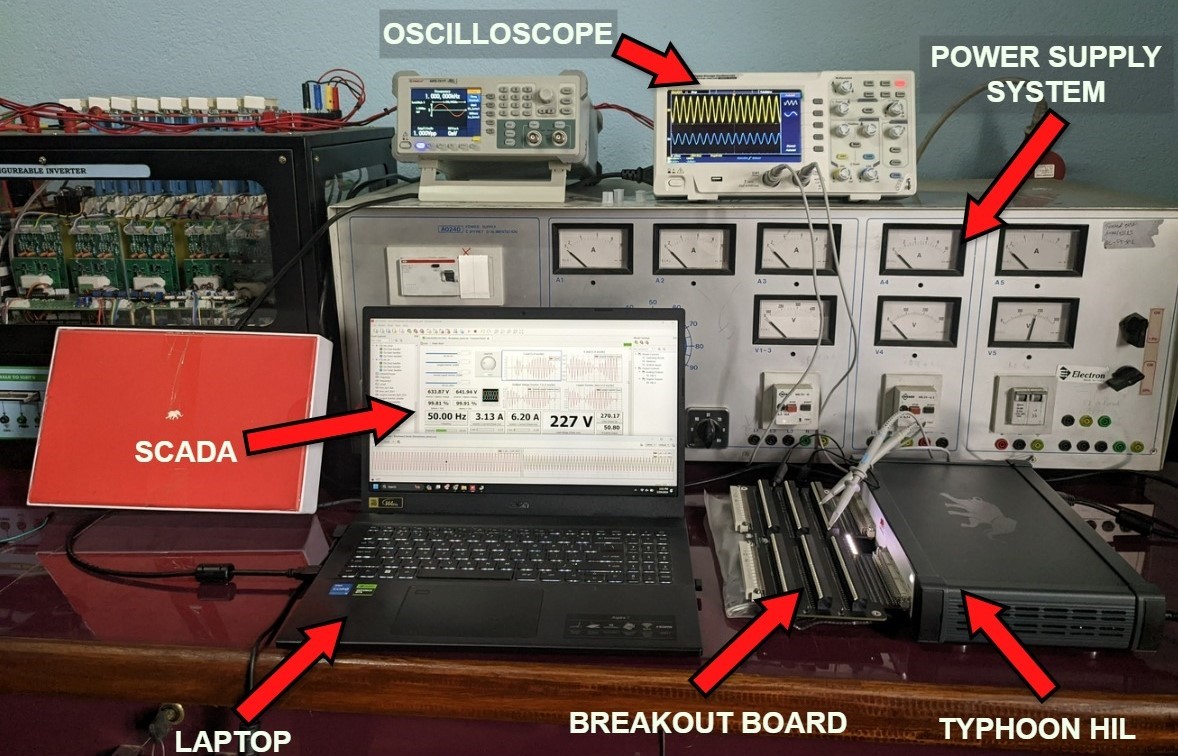}}
    \caption{Setup for Hardware-in-the-Loop: Typhoon HIL-404, Oscilloscope, Power Supply Unit and Virtual SCADA.}
    \label{hil_voc}
\end{figure}

The inverters are connected to the load via an RLC filter. The values of these filters depend on the ratio of their capacities for optimal power sharing. While this may not be necessary considering VRL, it is recommended to preserve the power-sharing characteristic. To validate the inverter system, a Hardware-in-the-Loop (HIL) setup using Typhoon HIL 404, Typhoon HIL SCADA and Oscilloscope is set up as seen in Figure \ref{hil_voc}. Due to the limitations of this setup, only two inverters were modelled. The oscilloscope is set up to show grid voltage and current, while the HIL-SCADA displays various other parameters: frequency, voltage, load and control loop values.

\begin{figure}
    \centering
    \includegraphics[width=0.45\textwidth]{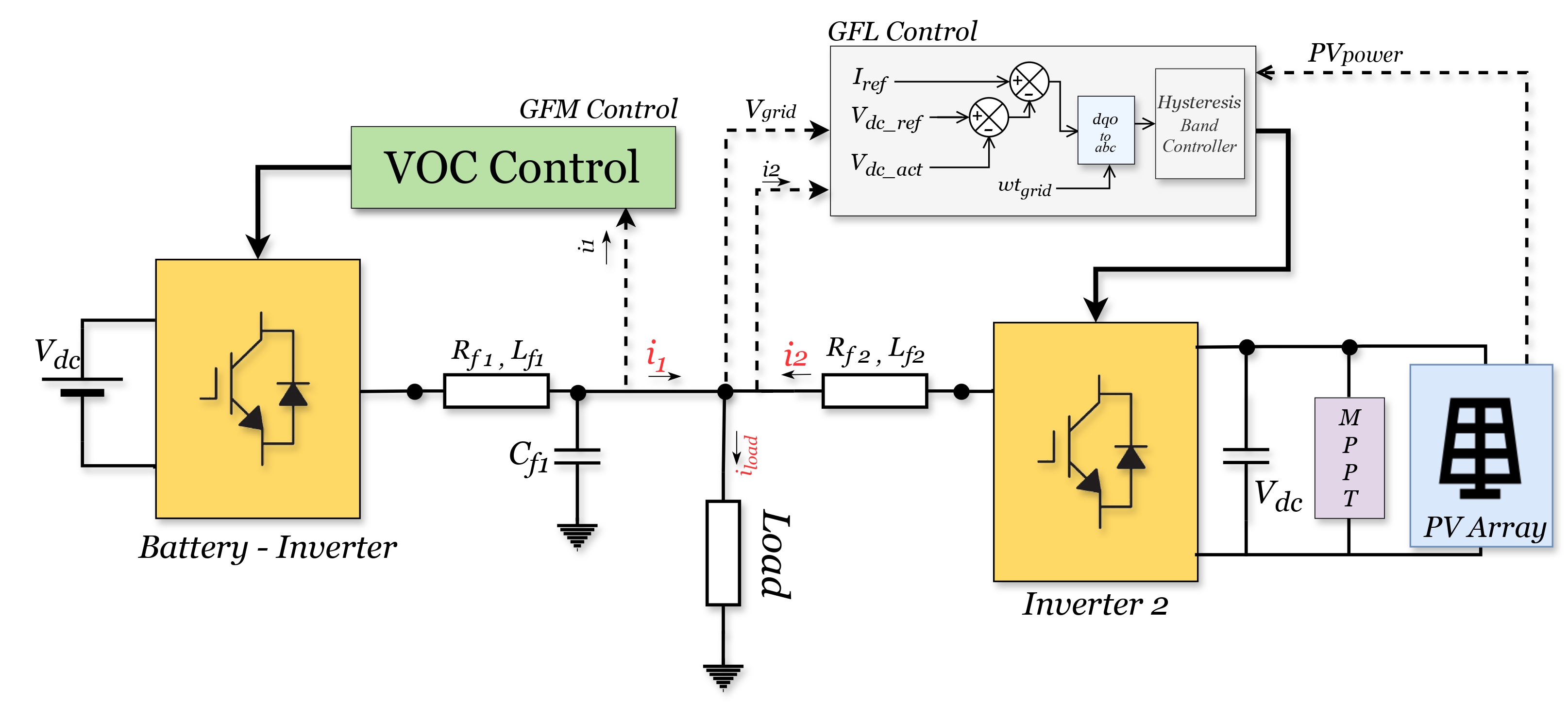}
    \caption{A battery-inverter system, with VOC, supplying a common load in parallel with a PV-inverter system, with hysteresis current control. }
    \label{voc_pv_grid}
\end{figure}

\subsection{PV Array}
A general PV-inverter system follows the grid, requiring a reference signal to regulate its output. The gate signals to each of the inverter switches are generated by a hysteresis current controller that uses relays to maintain the instantaneous value of its output current within a specified band around the reference. The magnitude of the output current is determined by the power generated by the PV array and is controlled to maintain the DC side voltage at a constant value of 600 V to obtain a 400 V line-line at the AC side.

\subsection{Micro-grid: GFM followed by Grid Following (GFL)}
The PV system operating in grid-following mode is connected in parallel with a battery-inverter source controlled by VOC logic, as seen in Figure \ref{voc_pv_grid}. This is done to replicate a small yet fully renewable grid with multiple sources. Here, the voltage and frequency of the grid are regulated by the battery-inverter source while PV injects its full generated power. Priority is given to the PV and only the remaining power mismatch is fulfilled by the battery source (provided it has the capacity). The stability of the grid is highly dependent on the capability of the battery-inverter source. Also, the load is set higher than the PV generation at all times.

\begin{figure}
  \centering
  \begin{minipage}[b]{0.247\textwidth}
    \includegraphics[width=\textwidth]{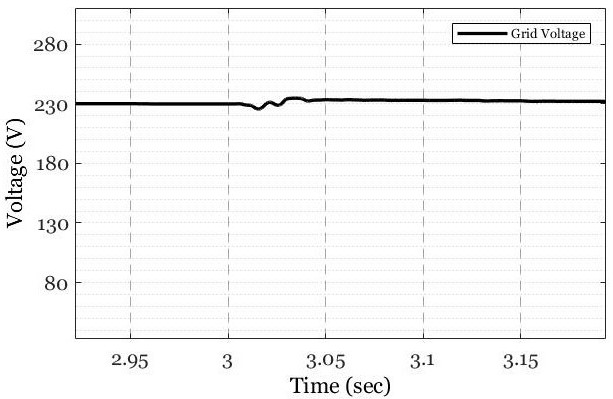}
    \caption{Voltage (RMS) profile during load decrease (at t = 3s).}
    \label{voc_voltage_graph}
  \end{minipage}
  \hfill
  \begin{minipage}[b]{0.228\textwidth}
    \includegraphics[width=\textwidth]{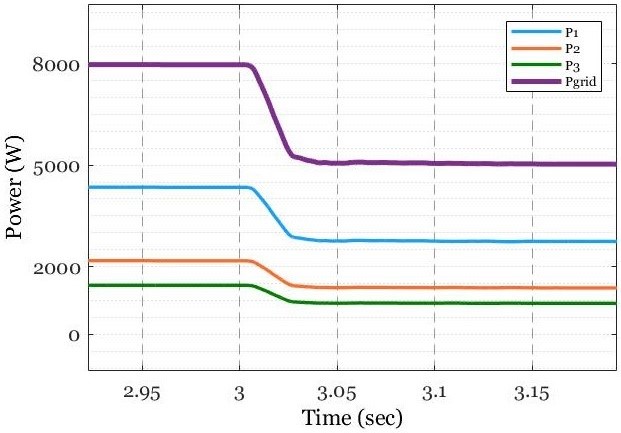}
    \caption{Power sharing in an appropriate ratio during steady operation and load change (at t = 3s).}
    \label{voc_power_graph}
  \end{minipage}
\end{figure}

\begin{figure}
    \centering
    \fbox{\includegraphics[width=0.45\textwidth]{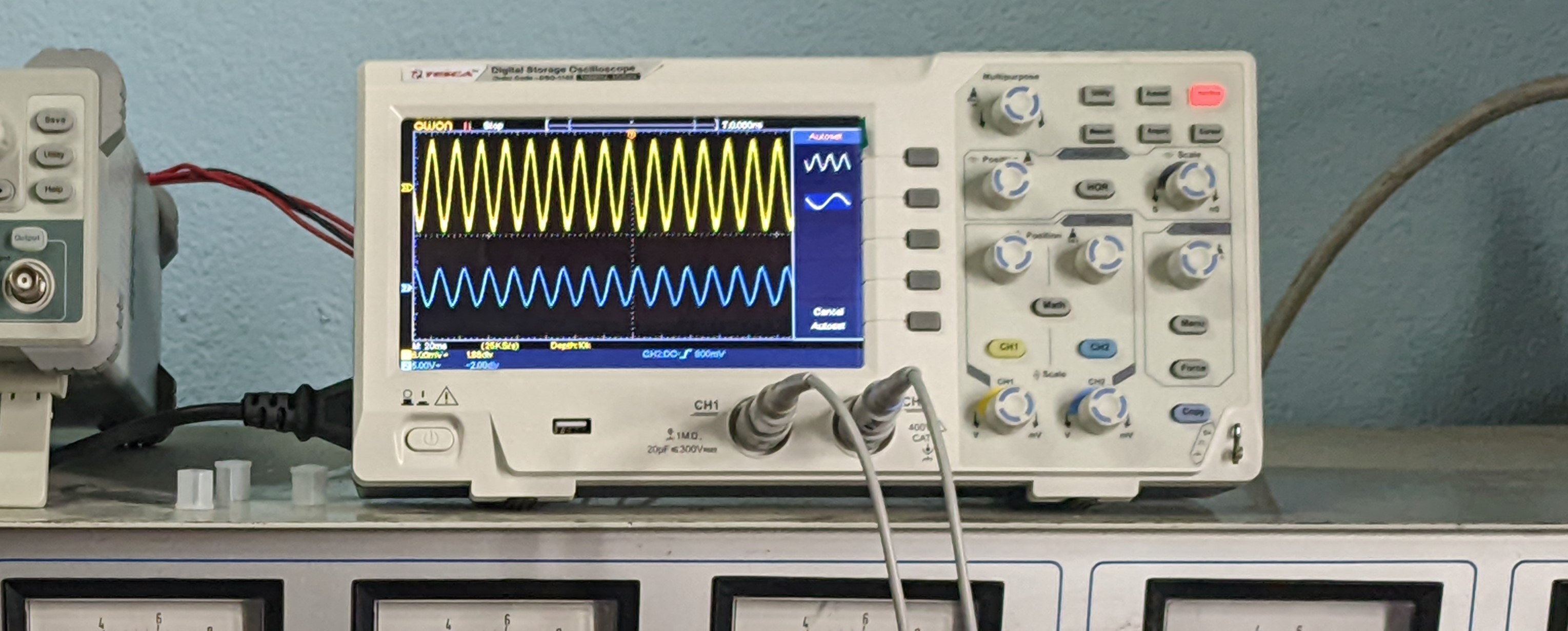}}
    \caption{Yellow: Voltage (maintained) and Blue: Current of the system immediately after a load change observed in the oscilloscope through real-time HIL simulation.}
    \label{hil_oscilloscope}
\end{figure}

\begin{figure}
  \centering
  \begin{minipage}[b]{0.238\textwidth}
    \includegraphics[width=\textwidth]{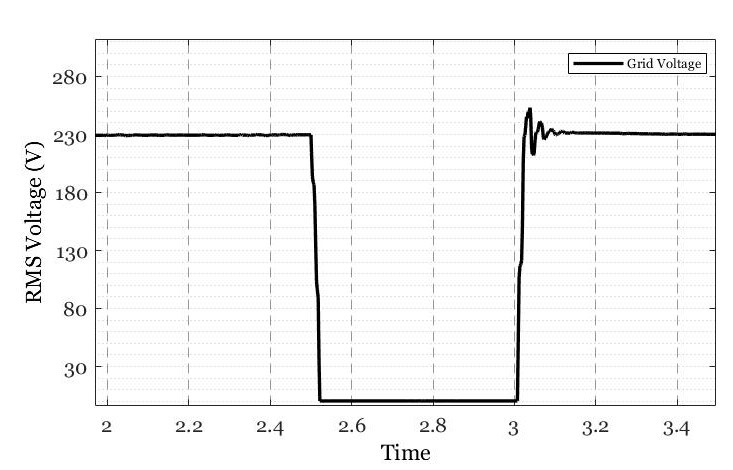}
    \caption{Voltage (RMS) profile before, during, and after a fault (t = 2.5s to 3s).}
    \label{voltage_fault}
  \end{minipage}
  \hfill
  \begin{minipage}[b]{0.238\textwidth}
    \includegraphics[width=\textwidth]{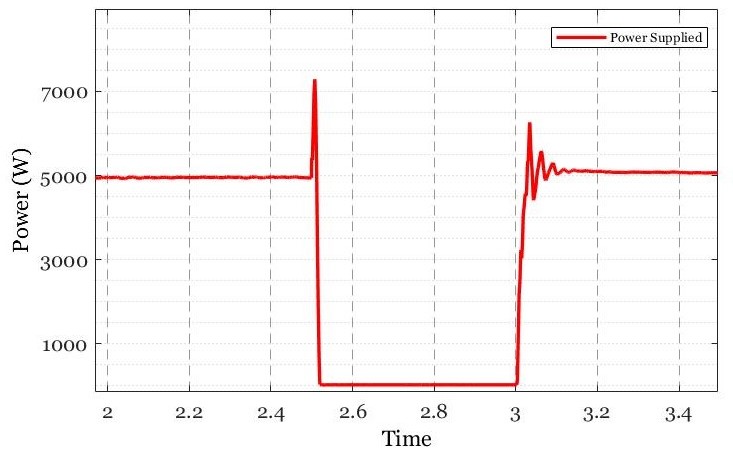}
    \caption{Total power supply before, during, and after a fault (t = 2.5s to 3s).}
    \label{power_fault}
  \end{minipage}
\end{figure}

\section{Results}
\subsection{VOC Microgrid}
The system as seen in Figure \ref{voc_grid} is supplying a load of 8kW at 230 V and 50 Hz. During a sudden load decrease by 3kW, at t = 3s, the voltage and power responses are seen in Figures \ref{voc_voltage_graph} and \ref{voc_power_graph} respectively. The VOC acts almost instantaneously, well within t = 3.05s, and the VRL loop maintains a constant voltage after the load change. During extreme conditions, such as faults, the controller has to be externally reset, to avoid slow responses by the PI controller. The working of the VOC in the HIL simulation is observed through the oscilloscope as seen in Figure \ref{hil_oscilloscope} which shows the voltage magnitude returning close to the reference after load changes, due to the VRL. The graphs in Figure \ref{voltage_fault} and \ref{power_fault} show the Root Mean Square (RMS) voltage and the total power supply in the grid before, during, and after a short circuit fault respectively. 

\subsection{PV and Battery Source Micro-grid}
\begin{figure}
  \centering
  \begin{minipage}[b]{0.235\textwidth}
    \includegraphics[width=\textwidth]{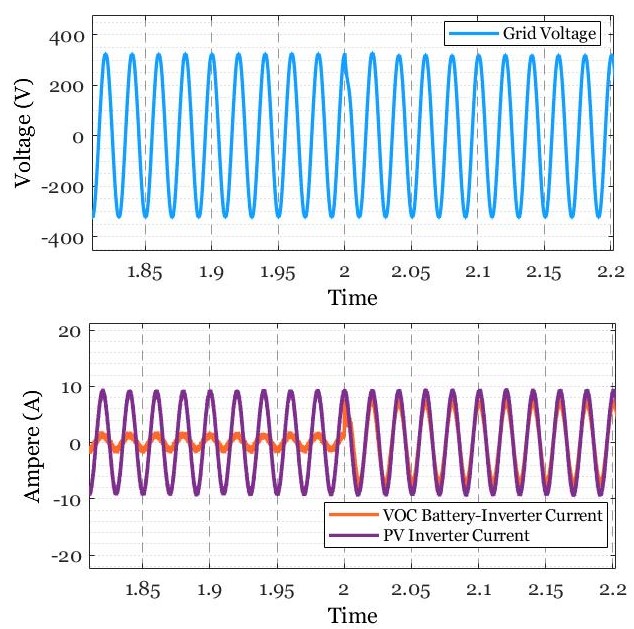}
    \caption{Voltage and Current profile of the PV plus Battery micro-grid system during a load increase at time = 2 seconds.}
    \label{voc_pv_voltage_current_graph}
  \end{minipage}
  \hfill
  \begin{minipage}[b]{0.235\textwidth}
    \includegraphics[width=\textwidth]{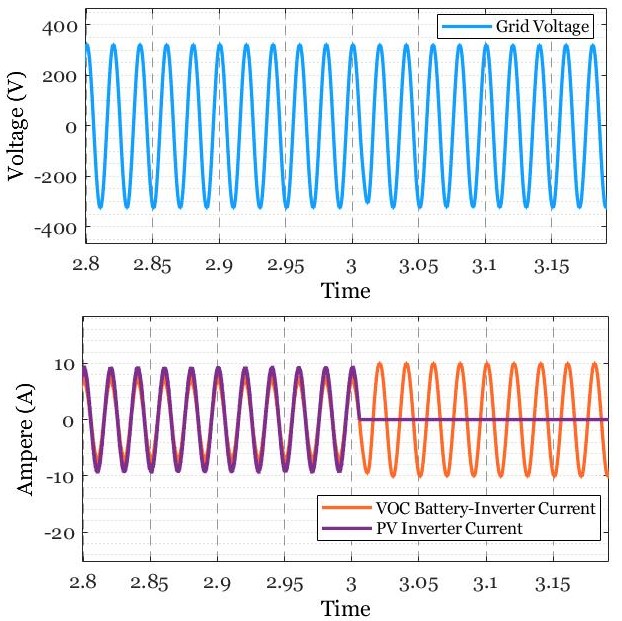}
    \caption{Voltage and Current profile of the PV plus Battery micro-grid system during disconnection of PV at time = 3 seconds.}
    \label{voc_pv_voltage_current_graph_decrease}
  \end{minipage}
\end{figure}

The system as seen in Figure \ref{voc_pv_grid} is supplying a load of 5kW initially. The PV source is generating at a capacity of 4.8kW ($>90\%$ PV penetration). System measurements, as seen in Figure \ref{voc_pv_voltage_current_graph}, show the voltage and current profiles during a sudden load increase of 3kW at 2 seconds. The additional load is supplied by the battery-inverter system since the PV array is set to inject all the power it generates. The voltage remains constant as the recovery action kicks in. The graph in Figure \ref{voc_pv_voltage_current_graph_decrease} shows the system parameters when the PV is suddenly disconnected (or it stops generating fully) at t = 3s. The battery-inverter system controlled by VOC will compensate for the demand mismatch instantly, depending on its capacity. The reaction time for current pick-up is instantaneous and the voltage regains its normal value at just over t = 3.05s.

\section{Conclusion}
Traditional inertia will no longer influence the stability of modern grids due to the increasing penetration of renewable and intermittent sources like PV and wind. Virtual oscillator control is an interesting approach for grid-forming inverters. This control methodology provides a quick initial response with inherent and accurate power sharing. Various architectures under the domain of VOC, like VdP, DZ, and AHO, have unique characteristics which fit them for particular situations. Voltage range limits may hinder the operating range of a VOC. In this paper, virtual oscillator control is used for inverter control along with a voltage recovery loop to surpass the restrictions due to voltage deviations. While the VRL can maintain voltage at a reference value, it requires extensive parameter tuning with an intricate reset-coordination system during anomalous events. A hysteresis-band-controlled current source inverter for a PV system is compatible with the DZ-VOC battery-inverter system. During load or generation fluctuations, the battery system acts immediately to compensate for any imbalances, provided it has the capacity. The results show an almost instantaneous response provided by the VOC control. Additionally, multiple VOC-operated inverters in parallel can share the load and synchronize with one another quickly and accurately without any external control. The robustness of VOC control, coupled with its numerous advantages will see significant use in the next generation of Grid Forming inverters.

%\bibliographystyle{IEEEtran} 
%\bibliography{bibliography}

\end{document}